\documentclass[pre,aps,twocolumn,showpacs]{revtex4}
\usepackage{amsmath,amssymb,graphicx,cases}
\usepackage{epsfig}
\usepackage{textcomp}

\begin{document}
\title{Survival of a static target in a gas of diffusing particles with exclusion}
\author{Baruch Meerson}
\affiliation{Racah Institute of Physics, Hebrew University of
Jerusalem, Jerusalem 91904, Israel}
\author{Arkady Vilenkin} \affiliation{Racah Institute of Physics, Hebrew University of
Jerusalem, Jerusalem 91904, Israel}
\author{P. L. Krapivsky} \affiliation{Department of Physics, Boston University, Boston,
  MA 02215, USA}


\begin{abstract}
Suppose that a lattice gas of constant density,
described by the symmetric simple exclusion process,  is brought in contact with a ``target": a spherical absorber of radius $R$. Employing the macroscopic fluctuation theory (MFT), we evaluate the probability ${\mathcal P}(T)$ that no gas particle hits the target until a long but finite time $T$. We also find the most likely gas density history  conditional on the non-hitting.  The results depend on the dimension of space $d$ and on the rescaled parameter $\ell=R/\sqrt{D_0T}$, where $D_0$ is the gas diffusivity. For small $\ell$ and $d>2$, ${\mathcal P}(T)$ is determined by an exact stationary solution of the MFT equations that we find. For large $\ell$, and for any $\ell$ in one dimension, the relevant MFT solutions are non-stationary. In this case $\ln {\mathcal P}(T)$ scales differently with relevant parameters, and it also depends on whether the initial condition is random or deterministic. The latter effects also occur if the lattice gas is composed of non-interacting random walkers. Finally, we extend the formalism to a whole class of diffusive gases of interacting particles.

\end{abstract}
\maketitle

\section{Statement of the problem}

Suppose that at $t=0$ a gas of diffusing particles of constant density $n_0$ is brought in contact with a spherical absorber of radius $R$ in $d$ dimensions.  The particles are absorbed upon hitting the absorber. Remarkably, this simple setting captures the essence of many diffusion-controlled chemical kinetic processes \cite{S16,C43,B60,BP77,R85,OTB89,book}. The evolution of the \emph{average} coarse-grained particle density of the gas is described by the diffusion equation
\begin{equation}
\partial_t n =\nabla \cdot [D(n) \nabla n],
\label{difeq}
\end{equation}
where $D(n)$ is the gas diffusivity. Here we will be interested in large fluctuations rather than in the average behavior. One important fluctuating quantity is the number of particles $N$ that is absorbed during a long time $T$. We will focus on two questions: (i) What is the probability that $N=0$, that is no particle hit the absorber until time $T$? (ii) What is the most likely history of the particle density of the gas conditional on the non-hitting until time $T$?

These questions also appear in the context of a search for an immobile target by a swarm of diffusing searchers, see e.g. Ref. \cite{Carlos} and references therein. This process has been studied extensively in the simplest case when the searchers are non-interacting random walkers (RWs). In this case $D(n)=D_0=\text{const}$, and the probability that the target survives until time $T$, ${\mathcal P}_{\text{RW}}(T)$ was found to exhibit the following long-time behavior \cite{ZKB83,T83,RK84,BZK84,BKZ86,BO87,Oshanin,BB03,BMS13}:
\begin{numcases}
{\!\! -\frac{\ln {\mathcal P}_{\text{RW}} (T)}{n_0}\simeq}
\frac{2 (D_0 T)^{1/2}}{\sqrt{\pi}}, \!\!\!\!\!\!   & $d=1$, \label{survivaldecay1}\\
\frac{4 \pi D_0 T}{\ln (D_0 T/R^2)},  \!\!\!\!\!\!   & $d=2$, \label{survivaldecay2}\\
(d-2) \Omega_d\,R^{d-2} D_0 T, \!\!\!\!\!\!& $d>2$, \label{survivaldecay3}
\end{numcases}
where $\Omega_d = 2\pi^{d/2}/\Gamma(d/2)$ is the surface area of the $d$-dimensional unit sphere,  and $\Gamma(z)$ is the gamma function.

Equations~(\ref{survivaldecay2}) and (\ref{survivaldecay3}) give the leading terms of the corresponding asymptotics at long times, when $\ell=R/\sqrt{D_0T}\ll 1$, i.e.,  the characteristic diffusion length $\sqrt{D_0T}$ is very large compared to the target radius $R$. Equation~(\ref{survivaldecay1}) is independent of $R$, and the parameter $\ell$ is irrelevant. As a result,  Eq.~(\ref{survivaldecay1}) becomes valid as soon as $T$ is much larger than the inverse microscopic hopping rate.

The target survival problem is a particular case of a more general problem of finding the complete statistics of particle absorption by the absorber. For the RWs, this problem has been recently studied in Ref. \cite{MR}.

\begin{figure}
\includegraphics[width=0.36\textwidth,clip=]{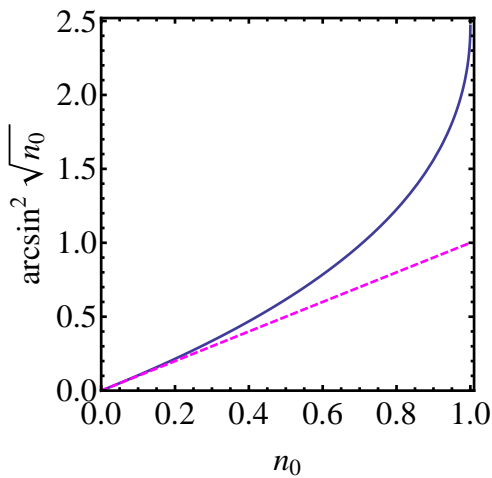}
\caption{(Color online) The function $\arcsin^2 \sqrt{n_0}$ (the solid line) which describes the density dependence of $-\ln {\cal P}(T)$ for the SSEP at $d>2$. The straight line shows  the corresponding density dependence for the gas of random walkers (RW).}
\label{arcsinsquared}
\end{figure}

Here we extend the target problem in several directions. First, we consider a lattice gas of \emph{interacting} searchers. Throughout most of the paper, we assume that the searchers interact via exclusion. This  can be a good simplistic model for studying diffusion-controlled chemical reactions in crowded environments such as a living cell \cite{crowded}. Specifically, we will consider a lattice gas described by the symmetric simple exclusion process (SSEP). In this process each particle can hop to a neighboring lattice site if that site is unoccupied by another particle. If it is occupied, the move is disallowed. The average behavior of this gas is still described by the diffusion equation (\ref{difeq}) with $D=D_0=\text{const}$ \cite{Spohn}, so the SSEP and the RWs are indistinguishable at the level of averages. However, as we show here, the long-time asymptotic of the target survival probability ${\mathcal P}(T)$  for the SSEP behaves differently from that for the RWs:
\begin{equation}\label{Actresultd}
-\ln {\mathcal P}\simeq (d-2) \Omega_d\,R^{d-2} D_0T \arcsin^2\sqrt{n_0},\;\; d>2.
\end{equation}
This expression has the same structure as Eq.~(\ref{survivaldecay3}), but it increases much faster with the gas density $n_0$  \cite{dimensions}; see Fig. \ref{arcsinsquared}.  We note that previous results for the SSEP
only included bounds on ${\cal P}$ \cite{SSEPqual}.

Second, we show that, for $d=1$,  the survival probability ${\mathcal P}(T)$ depends strongly on the initial condition. This effect does not require inter-particle interaction, and it also occurs for the RWs, as we show below. In particular, the asymptotic~(\ref{survivaldecay1}) is only valid after averaging over random initial distributions of particles, that is, for the annealed setting \cite{DG2009b,Evans}. We find a different result for a deterministic initial condition,
also called a quenched setting \cite{DG2009b,Evans}. For the RWs, the two results  for $\ln {\mathcal P}(T)$ differ by a numerical factor. For the SSEP, even their $n_0$-dependence is different for $d=1$.

Third, we demonstrate that the two basic one-dimensional solutions, annealed and quenched, play a central role in higher dimensions when one is interested in \emph{intermediate} asymptotics of ${\mathcal P}(T)$ for $\ell\gg 1$, that is when the diffusion length $\sqrt{D_0T}$ is much longer than the lattice constant, but much shorter than the absorber radius $R$.

Fourth, in addition to evaluating ${\mathcal P}(T)$ in different regimes, we also find the most likely history of the
gas density conditional on the target survival until time $T$. We achieve this result, and most of the others,  by employing
the macroscopic fluctuation theory (MFT) \cite{MFTreview}. This coarse-grained large-deviation formalism was unavailable when most of the studies of the target survival probability were performed. The MFT is well suited for the analysis of large deviations in lattice gases, including the (unlikely) target survival at long times that we are dealing with here. One of our central findings for the SSEP is that, in the long-time regime, $\ell \ll 1$, the most likely gas density profile for $d>2$ is, for most of the time, almost stationary:
\begin{equation}\label{qd}
    q(r) = \sin^2 \left[\left(1-\frac{\ell^{d-2}}{r^{d-2}}\right)\,\arcsin \sqrt{n_0}\right] ,
\end{equation}
where the radial coordinate is rescaled by the diffusion length $\sqrt{D_0T}$. It is this density profile that determines the leading term (\ref{Actresultd}) of the survival probability.

Last but not least, we extend our approach to a whole class of additional interacting diffusive gases.

In the next section we present the MFT formulation of the target survival problem. Section \ref{steady} deals with $d\geq 2$ for $\ell \ll 1$. Here
$\ln {\mathcal P}(T)$ is mostly contributed to by a \emph{stationary} solution of the MFT equations, independently of whether the setting is annealed or quenched.  We derive
these solutions, evaluate $\ln {\mathcal P}$ and verify the results for $d=3$ by solving the
MFT problem numerically. In Sec. \ref{nonstat} we study analytically and numerically the survival probability in non-stationary settings, deterministic and random, in all dimensions and at different densities. In Sec. \ref{extension} we extend our results for $d\geq 2$ to a broad class of interacting lattice gases.  Our main results are summarized in Sec. \ref{discussion}. In Appendix we present, for non-interacting RWs, exact microscopic derivations of ${\mathcal P}(T)$ for the annealed and quenched settings and for $d=1$, $2$ and $3$. Both the microscopic derivation and the MFT calculations show that, for $\ell\ll 1$,  the leading contribution to $\ln {\mathcal P}(T)$ is sensitive to the initial condition only in one dimension.

\section{Macroscopic fluctuation theory of target survival}
\label{MFT}

The macroscopic fluctuation theory (MFT) was developed for the analysis
of non-equilibrium steady states of diffusive lattice
gases~\cite{Bertini,Tailleur,D07,Bunin,HEPG}. Subsequently it was extended to
a host of non-stationary settings~\cite{DG2009b,Lecomte,KM_var,KMS,MS2013,MS2014,MR}.
The MFT, and its extensions to reacting particle systems~\cite{EK,MS2011},
have proven to be highly efficient and versatile.  Here we outline the MFT formulation, referring
the reader to the above references for further details.

The starting point for the derivation of the MFT can be a Langevin equation that provides a faithful large-scale description to a broad family of diffusive gases:
\begin{equation}
\label{Lang}
     \partial_t n = \nabla \cdot [D(n) \nabla n] +\nabla \cdot \left[\sqrt{\sigma(n)} \,\text{\boldmath$\eta$} (\mathbf{x},t)\right],
\end{equation}
where $\text{\boldmath$\eta$} (\mathbf{x},t)$ is a zero-average Gaussian noise, delta-correlated both in space and in time \cite{Spohn}. As one can see, a fluctuating diffusive gas is fully characterized by $D(n)$ and another coefficient, $\sigma(n)$, that comes from the shot noise and is equal to twice the mobility of the gas \cite{Spohn}. Essentially, the MFT formalism is a WKB theory (after Wentzel, Kramers and Brillouin) of the functional Fokker-Planck equation following from the Langevin equation (\ref{Lang}). The WKB theory employs, in a smart way, the typical number of particles in the relevant region of space as a large parameter \cite{Bertini,Tailleur,DG2009b,KMS}. In the MFT formalism, the particle number density field $q(\mathbf{x},t)$ and the canonically conjugate
``momentum" density field $p(\mathbf{x},t)$ obey the Hamilton equations
\begin{eqnarray}
  \partial_t q &=& \nabla \cdot \left[D(q) \nabla q-\sigma(q) \nabla p\right], \label{d1} \\
  \partial_t p &=& - D(q) \nabla^2 p-\frac{1}{2} \,\sigma^{\prime}(q) (\nabla p)^2, \label{d2}
\end{eqnarray}
where the prime denotes the derivative with respect to the argument.   Equations~\eqref{d1} and \eqref{d2} can be written in terms of
variational derivatives:
\begin{equation}
\partial_t q = \delta H/\delta p\,, \quad
\partial_t p = -\delta H/\delta q\,.
\end{equation}
Here
\begin{equation}
\label{Hamiltonian}
H[q(\mathbf{x},t),p(\mathbf{x},t)]= \int d\mathbf{x}\,\mathcal{H}
\end{equation}
is the Hamiltonian, and
\begin{equation}
\label{Ham}
\mathcal{H}(q,p) = -D(q) \nabla q\cdot \nabla p
+\frac{1}{2}\sigma(q)\!\left(\nabla p\right)^2
\end{equation}
is the Hamiltonian density. The spatial integration in Eq.~(\ref{Hamiltonian}), and everywhere in the following, is performed over the whole space outside the target. Because of the rotational symmetry of the problem, we assume that the solution only depends on the radial coordinate and time.  We will consider the target survival problem in an arbitrary dimension  $d$. The boundary conditions on the target are $q(r=R,t)=p(r=R,t)=0$ \cite{MR}, where the condition on $p(r=R,t)$ just fixes an arbitrary constant.  Far away from the target the gas is unperturbed, so  $q(r=\infty, t)=n_0$.  The boundary conditions in time are the following. At $t=0$ we prescribe
\begin{equation}\label{t0}
    q(r>R,t=0)=n_0,
\end{equation}
where, for the SSEP, $0<n_0<1$. This is a deterministic, or quenched, initial condition,
see Refs. \cite{DG2009b,Evans,KM_var,KMS,MR}. A random initial condition (that is, an annealed setting) is considered in Sec. \ref{app:annealed}.
Before focusing on the target survival problem, let us consider for a moment a slightly different setting where $N$,
the specified number of absorbed particles by time $t=T$, is arbitrary. This condition,
\begin{equation}\label{number}
\Omega_d \int_R^{\infty} dr\, r^{d-1}\,  [n_0-q(r,T)] = N,
\end{equation}
imposes an integral constraint on the solution. This constraint is identical to the one arising in the problem of statistics of integrated current during a specified time \cite{DG2009b,KM_var,MS2013,MS2014,MR}. A similar derivation yields the following boundary condition for $p$ at $t=T$:
\begin{equation}\label{pT}
    p(r,t=T)=\lambda \, \theta(r-R),
\end{equation}
where $\theta(\dots)$ is the Heaviside step function, and $\lambda$ is an a priori unknown Lagrange multiplier that is ultimately set by Eq.~(\ref{number}) \cite{DG2009b,MR}. Accordingly, we demand $p(r=\infty, t)=\lambda$. The particular case of $N=0$ in which we are interested here corresponds to $\lambda\to +\infty$ \cite{MR}. In this case
the total particle flux to the target vanishes at all times $0<t<T$.

The solution of the MFT equations for $q(r,t)$ yields the optimal trajectory: the most likely density history of the system conditional on the number of absorbed particles $N$.
Once $q(r,t)$ and $p(r,t)$ are found, we can calculate the mechanical action $S$ which yields $\ln {\mathcal P} (N)$
up to a pre-exponential factor:
\begin{eqnarray}
\label{actionmain}
  &-&\ln {\mathcal P} \simeq S = \Omega_d \int_0^T dt\, \int_R^{\infty} dr\, r^{d-1}\,\left(p\partial_t q-\mathcal{H}\right) \nonumber \\
  &=&\frac{1}{2}\,\Omega_d \,\int_0^T dt \int_{R}^{\infty} dr\,r^{d-1}\,
\sigma(q)\, (\partial_r p)^2. 
\end{eqnarray}
For the SSEP $D(q)=D_0=\text{const}$ and $\sigma(q)=2 D_0 q(1-q)$ \cite{Spohn}, and
Eq.~(\ref{actionmain}) becomes
\begin{equation}\label{actionmainSSEP}
  -\ln {\mathcal P} \simeq S = \Omega_d D_0\,\int_0^T dt \int_{R}^{\infty} dr\,r^{d-1}\,
q(1-q)\, (\partial_r p)^2.
\end{equation}
Upon rescaling $t$ by $T$ and $r$ by $\sqrt{D_0T}$ \cite{footnoteD}, we can effectively put $T=1$ in Eqs.~(\ref{number}) and (\ref{pT}) and replace $R$ by $\ell =R/\sqrt{D_0T}$ and $N$ by $\nu=N/(D_0T)^{d/2}$ everywhere. Equation~(\ref{actionmainSSEP}) for the SSEP becomes
\begin{equation}
\label{actionscaled1}
-\ln {\cal P} \simeq (D_0T)^{d/2} s(\ell,\nu,n_0),
\end{equation}
where
\begin{equation}
\label{actionscaled2}
s= \Omega_d \int_0^1 dt \int_{\ell}^{\infty} dr\,r^{d-1}\, q(1-q)\, (\partial_r p)^2.
\end{equation}
We are interested in the limit of $s(\ell,\nu,n_0)$ as $\nu\to 0$.
In one spatial dimension, $d=1$, the parameter $R$ (and hence $\ell=R/\sqrt{D_0 T}$) is irrelevant because of the translational symmetry of the ensuing MFT problem. We will consider this case in Sec. \ref{1d}.
For $d\geq 2$ there are two natural limiting cases: of small and large $\ell$.

\section{$\ell\ll 1$: quasi-stationary fluctuations}
\label{steady}

\subsection{$d>2$}
\label{larged}

\begin{figure}
\includegraphics[width=0.4\textwidth,clip=]{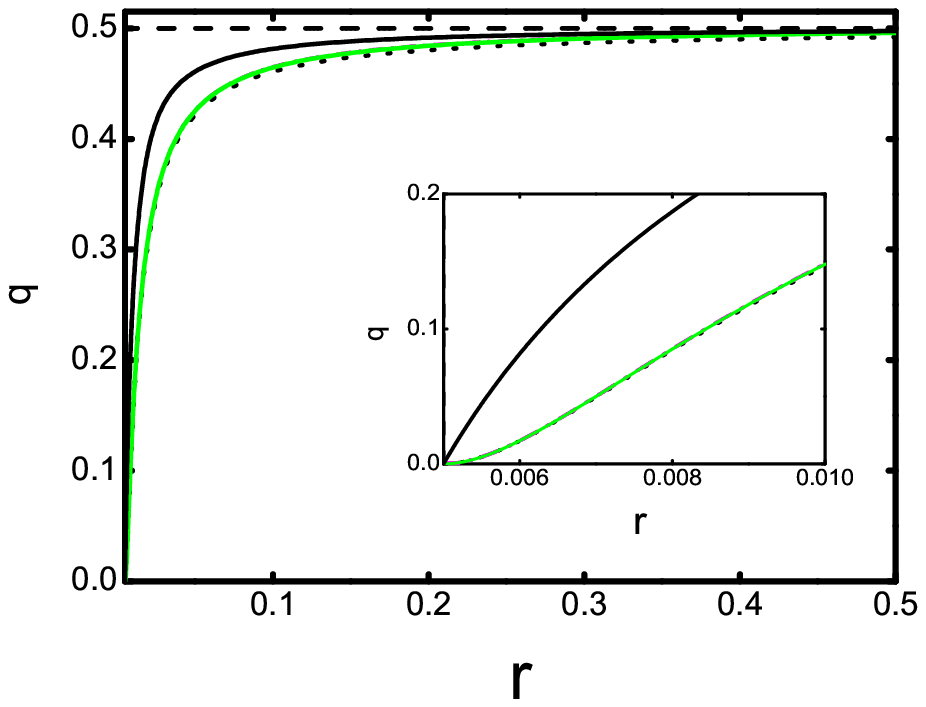}
\includegraphics[width=0.4\textwidth,clip=]{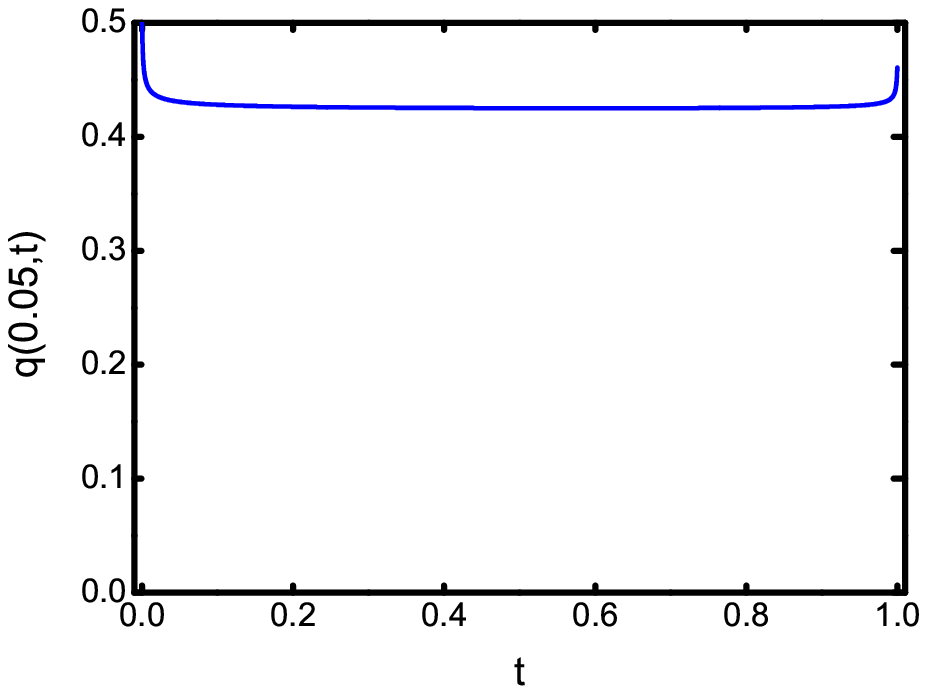}
\caption{(Color online) Upper panel: Theoretical stationary density profile for the SSEP at $d=3$, Eq.~(\ref{q3}) (dots), and numerical density profiles for $t=0.25$ and $t=0.75$ (indistinguishable), and
for $t=1$. The (deterministic) initial condition for the numerical solution is shown by the dashed line. Inset: a blowup of the region close to $r=\ell$. The numerical procedure is described in Sec. \ref{numeric1d}. Lower panel: $q(r=0.05,t)$ found numerically; the boundary layers at $t=0$ and $t=1$ are clearly seen. The parameters are $\ell=5 \cdot 10^{-3}$ and $n_0=0.5$. For these parameters, the theoretical rescaled action from Eq.~(\ref{s3d}) is $s\simeq 3.876 \cdot 10^{-2}$. The rescaled
action found numerically from Eq.~(\ref{actionscaled2}) is $3.916\cdot 10^{-2}$. The coordinate $r$ is rescaled by $\sqrt{D_0T}$, and time is rescaled by $T$. }  
\label{3dthnum}
\end{figure}

A small $\ell$ in the \emph{deterministic} theory, described by Eq.~(\ref{difeq}), means that $T$ is much longer than the characteristic diffusion time $R^2/D_0$ needed for the gas density to approach a steady state around the target. As a result, the average particle flux to the target can be determined by using the \emph{stationary} solution of the diffusion equation. For $D(n)=D_0=\text{const}$ this reduces to solving the Laplace equation $\nabla^2 n=0$ with the boundary conditions $n(r=R)=0$ and $n(r=\infty)=n_0$, leading to
\begin{equation}\label{MFsteady}
    n(r)=n_0\left(1-\frac{R^{d-2}}{r^{d-2}}\right),\;\;\;\;\;d>2.
\end{equation}
We argue that same logic holds for \emph{fluctuations}, including those responsible for the survival probability. Hence, when $\ell \ll 1$, the leading order contribution to the action $S$ from Eq.~(\ref{actionmain}) comes from the \emph{stationary} solution of the MFT equations that obeys the boundary conditions in space, but not the boundary conditions in time. For such solutions Eqs.~(\ref{d1}) and (\ref{d2}) become
\begin{eqnarray}
  &&r^{d-1}\left[-D(q) \frac{dq}{dr} +\sigma(q) v\right]= j= \text{const}, \label{zeroflux}\\
  &&\frac{D(q)}{r^{d-1}} \,\frac{d}{dr}\left(r^{d-1}  v\right)+\frac{1}{2}\,\sigma^{\prime}(q) v^2= 0, \label{steadyv}
\end{eqnarray}
where $v(r)\equiv dp/dr$.
The target survival implies that the particle flux at $r=\ell$ vanishes at all times $0<t<T$. Therefore $j=0$, and from Eq.~(\ref{zeroflux}) $v=(D/\sigma) (dq/dr)$. Plugging this into Eq.~(\ref{steadyv}) we obtain
\begin{equation}
\label{steadyeqgen}
\nabla^2_r q+\left(\frac{D^{\prime}}{D}-\frac{\sigma^{\prime}}{2\sigma} \right) \left(\frac{dq}{dr}\right)^2=0,
\end{equation}
where
$$
\nabla^2_r = \frac{1}{r^{d-1}}\,\frac{d}{dr}\left(r^{d-1}\,\frac{d}{dr} \right)
$$
is the spherically symmetric Laplace operator in $d$ dimensions. For the SSEP Eq.~(\ref{steadyeqgen}) reads
\begin{equation}\label{steadySSEP}
\nabla^2_r q+\frac{2q-1}{2q(1-q)}  \left(\frac{dq}{dr}\right)^2=0.
\end{equation}
Remarkably, the substitution $q(r)=\sin^2 u (r)$ reduces the nonlinear ordinary differential equation (\ref{steadySSEP}) to the spherically symmetric Laplace equation in $d$ dimensions:
\begin{equation}\label{Laplace}
\nabla^2_r u = 0.
\end{equation}
The boundary conditions $q(\ell)=0$ and $q(\infty)=n_0$ become $u(\ell)=0$ and $u(\infty)=\arcsin \sqrt{n_0}$. Solving this problem and returning to $q$,  we obtain Eq.~(\ref{qd}). This is the most likely density profile conditional on survival of the target until time $T$.  Now we can calculate $v(r)$:
\begin{eqnarray}\label{vd}
    v(r)&=&\frac{1}{2q(1-q)}\frac{dq}{dr} \nonumber \\
        &=&\frac{2(d-2) \ell^{d-2}\,\arcsin\sqrt{n_0}}{r^{d-1} \sin\left[2  \left(1-\frac{\ell^{d-2}}{r^{d-2}}\right)\,\arcsin\sqrt{n_0}\right]}.
\end{eqnarray}
In particular, for $d=3$
\begin{eqnarray}
  q(r) &=& \sin^2 \left[\left(1-\frac{\ell}{r}\right)\,\arcsin \sqrt{n_0}\right], \label{q3}\\
  v(r) &=& \frac{2\ell \arcsin\sqrt{n_0}}{r^2 \sin\left[2 \left(1-\frac{\ell}{r}\right)\,\arcsin \sqrt{n_0}\right]}.
  \label{v3}
\end{eqnarray}
The asymptotic of $q(r)$
near the target,
\begin{equation}
\label{qquadr}
    q(r-\ell\ll \ell)\simeq (d-2)^2 \arcsin^2(\sqrt{n_0})\,\left(\frac{r}{\ell}-1\right)^2,
\end{equation}
is quadratic in $r-\ell$. Also notable is a diverging asymptotic
of $v(r)=dp/dr$ near the target:
\begin{equation}\label{vdiverge}
    v(r-\ell\ll \ell) \simeq \frac{1}{r-\ell}
\end{equation}
which is independent of $n_0$. The asymptotic behaviors near the target assure that the particle flux to the target vanishes. Furthermore,
each of the two terms in the flux, see Eq.~(\ref{zeroflux}), vanish separately. As it turns out, these features,
including the 'one over the distance' asymptotic (\ref{vdiverge}), are quite universal: they are observed, for $0\leq t<1$, in the quenched and annealed settings and in all dimensions (including $d=1$ where the MFT solution is \emph{non}-stationary) for all lattice gases that behave as non-interacting RWs at low densities.   
An example of the stationary gas density profile for $d=3$ is shown in Fig. \ref{3dthnum}.

In spite of the singularity of $v(r)$ at $r=\ell$, the action~(\ref{actionscaled2}) is bounded, and we obtain
\begin{equation}
\label{actresultd}
s= (d-2)\, \Omega_d\, \ell^{d-2} \arcsin^2\sqrt{n_0},\;\;\;\; d>2,
\end{equation}
and arrive at Eq.~(\ref{Actresultd}). In particular, for $d=3$
\begin{equation}\label{s3d}
s = 4 \pi \ell \arcsin^2\sqrt{n_0}
\end{equation}
and
\begin{equation}
\label{S3d}
-\ln {\mathcal P}\simeq S =4 \pi \,R D_0 T\,\arcsin^2\sqrt{n_0}.
\end{equation}
Notice that as $n_0$ approaches $1$, the asymptotic survival probability goes down rapidly but remains non-zero.

As the solution (\ref{qd}) and (\ref{vd}) is stationary, the survival probability is independent, in the leading order, of whether the particles are distributed randomly or deterministically at $t=0$. Here for very long times, $D_0T\gg R^2$, the optimal fluctuation becomes unconstrained by the process duration, and details of the initial condition become irrelevant. As we will see in Sec. \ref{nonstat}, the situation changes for $d=1$, and for any $d$ when $\ell\ll 1$.

For $n_0\ll 1$, Eq.~(\ref{S3d}) reduces to Eq.~(\ref{survivaldecay3}) for the RWs. Further, Eqs.~(\ref{qd}) and (\ref{vd}) become
\begin{eqnarray}
    q(r) &=& n_0 \left(1-\frac{\ell^{d-2}}{r^{d-2}}\right)^2,\label{qdRW} \\
    v(r) &=& \frac{d-2}{r} \left[\left(\frac{r}{\ell}\right)^{d-2}-1\right]^{-1}. \label{vdRW}
\end{eqnarray}
These low-density asymptotics for the SSEP represent exact solutions for the RWs, where $D(q)=D_0=\text{const}$
and $\sigma(q)=2 D_0 q$ \cite{Spohn}.

The stationary solution (\ref{qd}) and (\ref{vd}), or (\ref{qdRW}) and (\ref{vdRW}), does not satisfy the boundary conditions in time. To accommodate
these boundary conditions, the full time-dependent solutions of the MFT problem must develop narrow boundary layers in time at $t=0$ and $t=1$,  cf. Ref. \cite{MVS}. The boundary layers only give a subleading contribution to $s$.  We verified this scenario numerically for the SSEP. Figure \ref{3dthnum} shows the density history obtained by solving the full non-stationary MFT problem numerically for a sufficiently small $\ell$ and $d=3$. The numerical algorithm is described in Sec. \ref{numeric1d}. One can see that $q(x,t)$ stays almost constant for most of the time and, at these times, agrees very well with the theoretical prediction (\ref{q3}). The rescaled action, found numerically from Eq.~(\ref{actionscaled2}),  is also close to the theoretical prediction, Eq.~(\ref{s3d}).

\subsection{$d=2$}
\label{2d}

In the marginal case of $d=2$ logarithmic corrections appear. For $d=2$ all non-constant solutions of the circularly-symmetric Laplace's equation (\ref{Laplace}) diverge logarithmically with $r$. As a result,
a reasonable stationary solution of Eq.~(\ref{steadySSEP}) can only be obtained if we introduce a cutoff distance $L$
(to remind the reader, all lengths are rescaled by $\sqrt{D_0T}$):
\begin{equation}
\label{q2}
q(r) =
\begin{cases}
\sin^2 \left(\frac{\ln \frac{r}{\ell} \arcsin \sqrt{n_0}}{\ln \frac{L}{\ell}}\right),   & \ell\leq r \leq L,\\
n_0,  & r>L.
\end{cases}
\end{equation}
Correspondingly,
\begin{equation}
\label{v2}
v(r) =
\begin{cases}
\frac{2 \arcsin \sqrt{n_0}}{r \ln \frac{L}{\ell}\,\sin\left(\frac{2 \ln\frac{r}{\ell} \arcsin \sqrt{n_0}}{\ln \frac{L}{\ell}}\right)},   & \ell\leq r \leq L,\\
0,  & r>L.
\end{cases}
\end{equation}
Although there is a derivative jump in $q(r)$ at $r=L$, and divergence of $v(r)$ at $r=\ell$ and $r=L$, the flux is continuous (and equal to zero) everywhere.  The cutoff $L$ in these formulas should be chosen ${\mathcal O} (1)$: in the original variables it is of the order of the characteristic diffusion length $\sqrt{D_0T}$ where the stationary solution crosses over to a time-dependent one. The uncertainty the cutoff introduces only affects the argument of the logarithm. As a result, Eqs.~(\ref{q2}) and (\ref{v2}) are correct with logarithmic accuracy. The same happens  \cite{Carslaw} if one circumvents the tedious exact time-dependent solution of the two-dimensional ($2D$) diffusion equation (\ref{difeq}) and solves instead the Laplace equation $\nabla^2 n=0$ for the purpose of computing the average particle flux onto the absorber.


The rescaled action is $s\simeq 2 \pi\,[\ln(1/\ell)]^{-1}\arcsin^2\! \sqrt{n_0}$
with the same logarithmic accuracy. Thus
\begin{equation}\label{action2}
-\ln {\mathcal P}\simeq S\simeq \frac{2 \pi D_0 T\,\arcsin^2 \sqrt{n_0}}{\ln\frac{\sqrt{D_0T}}{R}},\;\;\;\;\;\;R\ll \sqrt{D_0T}.
\end{equation}
For $n_0\ll 1$, Eq.~(\ref{action2}) reduces to Eq.~(\ref{survivaldecay2}) as expected.

\section{Non-stationary fluctuations}
\label{nonstat}

In the short-time limit, $\ell\gg 1$, the deterministic theory of diffusion-controlled absorption is non-stationary.
For $d=1$ the non-stationarity holds, in the deterministic theory, for any $\ell$. Again, we argue that the same features hold in the context of survival probability.

\subsection{$d=1$, deterministic initial condition}
\label{1d}

For the SSEP in one dimension Eqs.~(\ref{d1}) and (\ref{d2}) can be written as
\begin{eqnarray}
  \partial_t q &=& \partial_{x}^2 q-2\partial_x \left[q(1-q) \partial_x p\right], \label{qt} \\
  \partial_t p &=& - \partial_{x}^2 p+(2 q-1)(\partial_x p)^2, \label{pt}
\end{eqnarray}
whereas the Hamiltonian density~(\ref{Ham}) becomes
\begin{equation}
\label{w}
    {\mathcal H}(q,p) =-\partial_x p \,\partial_x q+q(1-q)(\partial_x p)^2.
\end{equation}
Here, and in most of the following exposition on the SSEP and RW, we put $D_0=1$. We will consider a one-sided problem and put the absorbing wall at $x=0$, so that $q(x=0,t)=p(x=0,t)=0$.
We assume a deterministic initial condition,
\begin{equation}\label{t01}
    q(x>0,t=0)=n_0,  \;\;\;0<n_0<1,
\end{equation}
and demand $q(x=\infty,t)=n_0$.  Upon rescaling $t$ by $T$ and $x$ by $\sqrt{T}$ Eq.~(\ref{pT}) becomes
\begin{equation}\label{pT1}
    p(x,t=1)=\lambda\, \theta(x),
\end{equation}
and we also have $p(x=\infty,t)=\lambda$. We remind the reader that $\lambda$ is ultimately set by the number of absorbed particles: when this number goes to zero,  $\lambda \to \infty$ \cite{MR}.

Once $q(x,t)$ and $p(x,t)$ are found, we obtain
\begin{eqnarray}
-\ln {\mathcal P}  &\simeq &  \sqrt{T}\, s_1(n_0), \label{1dresult}\\
  s_1(n_0) &=& \int_0^1 dt \int_{0}^{\infty} dx\,
q(1-q)\, (\partial_x p)^2, \label{actionscaled1d}
\end{eqnarray}
where the subscript in $s_1$ refers to $d=1$.

We have been unable to solve this problem exactly for arbitrary $n_0$. In the following we solve it in the limit of $n_0\ll 1$, when the SSEP reduces to RWs. Based on these results, we then compute the next-order correction in $n_0$ perturbatively. At the end of this subsection we solve the problem numerically for a range of values of $n_0$.

\subsubsection{Low-density limit: Non-interacting random walkers}
\label{RW}

\begin{figure}
\includegraphics[width=0.35\textwidth,clip=]{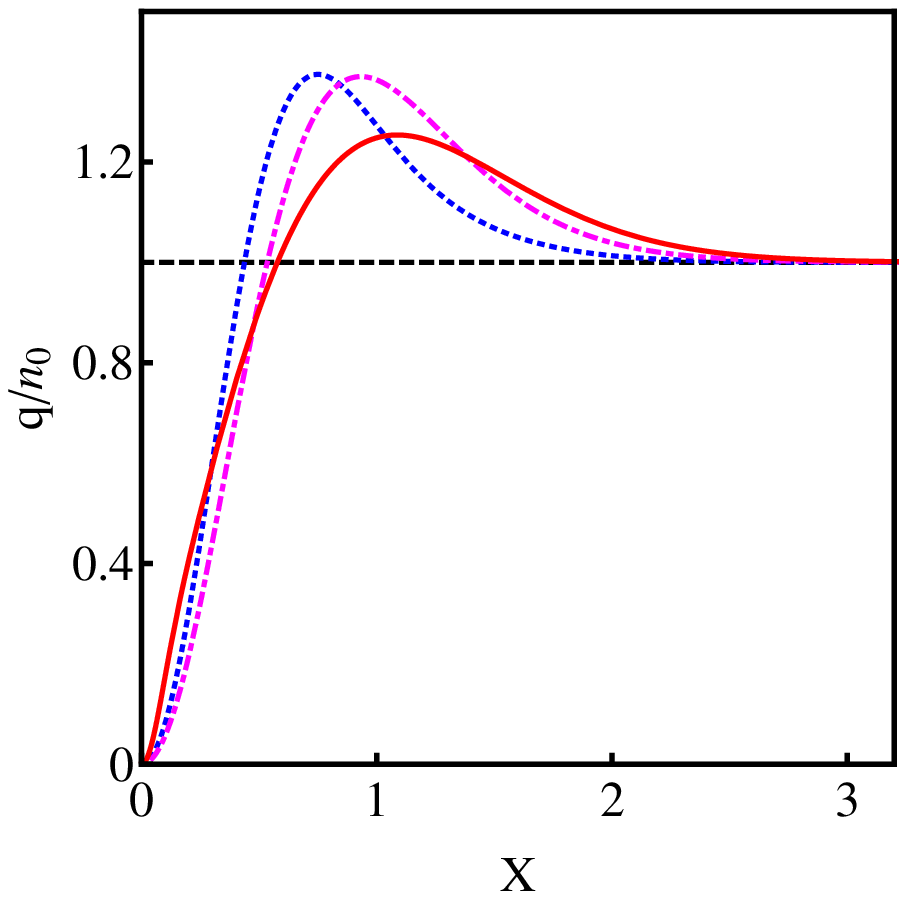}
\includegraphics[width=0.35\textwidth,clip=]{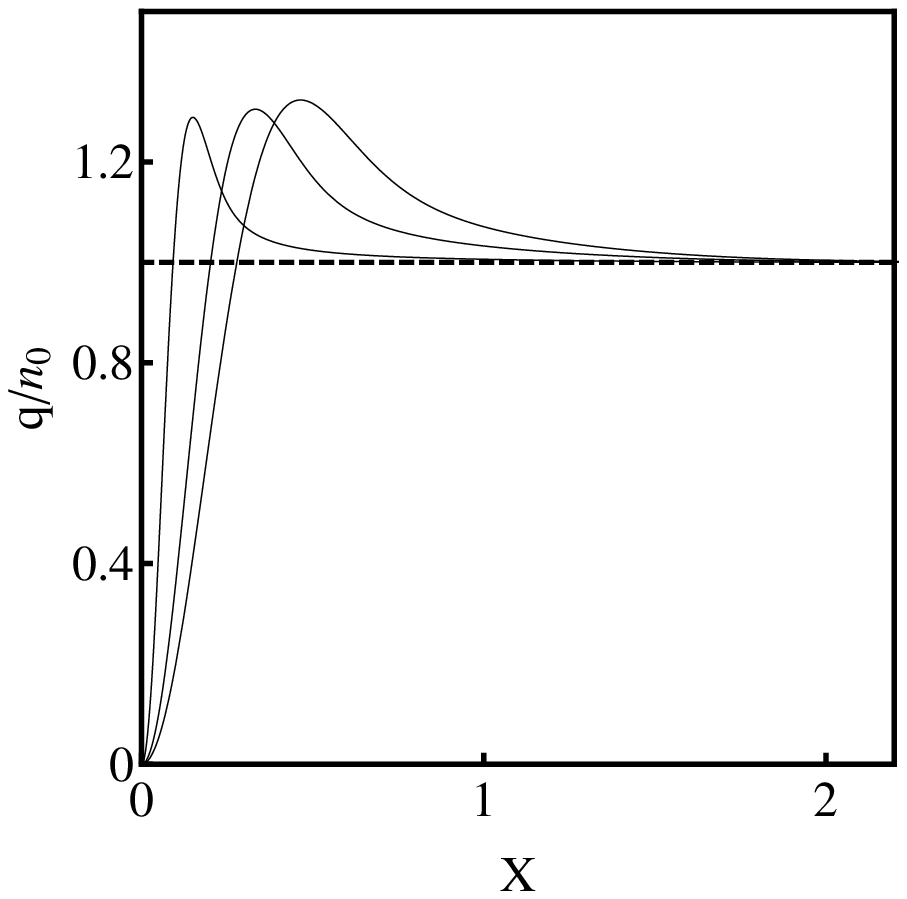}
\caption{(Color online) The most likely density history of the RWs at $d=1$, conditional on the zero flux to the target (located at $x=0$) for a deterministic initial condition with density $n_0$. Shown is $q/n_0$ from Eq.~(\ref{qRW}) versus $X=x/2$ where $x$ is rescaled by $\sqrt{T}$ (here $D_0=1$). Upper panel:  $t=0$ (dashed line), $1/3$ (dash-dotted line), $2/3$ (dotted line) and $1$ (solid line). Lower panel:  $t=0$,
$0.01$, $0.05$ and $0.1$. Time is rescaled by $T$.}
\label{RWquenchedq}
\end{figure}

In the limit of $n_0\ll 1$ we can drop $h_1 = - q^2 (\partial_x p)^2$ in the Hamiltonian density (\ref{w}), and the corresponding terms in the MFT equations, arriving at the RW model. As in other examples \cite{DG2009b,KMS,MR}, the MFT problem for the RW is solvable by the Hopf-Cole transformation $Q = q e^{-p}$ and $P = e^p$. This is because, in the new variables, the Hamilton equations are decoupled:
\begin{eqnarray}
  \partial_tQ &=& \partial_x^2 Q\,, \label{Qt}\\
   \partial_tP &=& - \partial_x^2 P\,. \label{Pt}
\end{eqnarray}
We can solve the anti-diffusion equation (\ref{Pt}) backward in
time, with the initial condition $P(x,T)=1+(e^{\lambda}-1) \theta(x)$ and the
boundary conditions $P(0,t)=1$ and $P(\infty,t)=e^{\lambda}$.  The solution
is
\begin{equation}\label{stepP}
    P(x,t)=1+(e^{\lambda}-1)\, \text{erf}\left(\frac{X}{\sqrt{1-t}}\right),
\end{equation}
where $X=x/2$, and $\text{erf}\, z =(2/\sqrt{\pi}) \int_0^z e^{-u^2} du$ is the error function.   At $t=0$
we obtain
\begin{equation*}
Q(x,0)=\frac{q(x,0)}{P(x,0)}=\frac{n_0}{1+(e^{\lambda}-1)\,\text{erf}\,X}.
\end{equation*}
This expression is the initial condition for the diffusion
equation~(\ref{Qt}) forward in time. The boundary conditions are
$Q(0,t)=q(0,t)/P(0,t)=0$ and $Q(\infty,t)=0$.  The solution is
\begin{equation}
  Q(x,t)=\frac{1}{\sqrt{\pi t}} \int_0^{\infty} d\mu\,\frac{e^{-\frac{(X-\mu )^2}{t}}-e^{-\frac{(X+\mu)^2}{
   t}}}{1+\left(e^{\lambda }-1\right) \text{erf} \,\mu}. \label{QRW}
\end{equation}
Transforming back to $q$ and $p$, and taking the limit of $\lambda \to \infty$ \cite{MR}, we obtain
\begin{eqnarray}
  q(x,t) &=& \frac{n_0}{\sqrt{\pi t}}\,\text{erf}\left(\frac{X}{\sqrt{1-t}}\right) \nonumber \\
    &\times&  \int_0^{\infty} d \mu \,\frac{e^{-\frac{(X-\mu)^2}{t}}-e^{-\frac{(X+\mu)^2}{t}}}{\text{erf}\,\mu} \,, \label{qRW}\\
  v(x,t)  &=&  \partial_x p(x,t) = \frac{e^{-\frac{X^2}{1-t}}}{\sqrt{\pi(1-t)}\,
   \text{erf} \left(\frac{X}{\sqrt{1-t}}\right)}, \label{vRW}
\end{eqnarray}
Figure \ref{RWquenchedq} depicts the density history of the system as described by Eq.~(\ref{qRW}). The lower panel shows a density ``void" that forms immediately. Also noteworthy is a density peak that accompanies the void formation. To the right of the density peak $v$ is very small,
and the dynamics is essentially governed by the deterministic equation~(\ref{difeq}) and corresponds to a diffusive outflow of the gas.  At $t=1$ we obtain
\begin{equation}\label{qt1}
 q(x,1) = \frac{n_0}{\sqrt{\pi}}\,\int_0^{\infty} d \mu \,\frac{e^{-(X-\mu)^2}-e^{-(X+\mu)^2}}{{\text{erf}\,\mu}} \,,
\end{equation}
As one can see, $q(x,1)$ behaves linearly in $x$ at small $x$. At $0<t<1$, however, $q(x,1)$ is quadratic at small $x$,
as in the stationary solution derived above.  Now,  $v(x,t)$, as described by Eq.~(\ref{vRW}), again exhibits the universal 'one over the distance' asymptotic. Indeed,  at $x\ll \sqrt{1-t}$,
\begin{equation}\label{sing1d}
    v\simeq \frac{1}{x},
\end{equation}
independent of time. This asymptotic already holds at $t=0$. The character of singularity at $x=0$ only changes at $t=1$, as $v(x,t=1)$ is equal to $\delta(x)$ with an infinite prefactor $\lambda \to \infty$.

To compute the rescaled action $s_1^{\text{RW}}$ we write
\begin{equation}\label{actionRW1}
    s_1^{\text{RW}}=\int_0^1 dt \int_{0}^{\infty} dx\,q(x,t)\, v(x,t)^2,
\end{equation}
which follows from Eq.~(\ref{actionscaled1d}) at $q\ll 1$. It is more convenient, however, to use the formula (derived in \cite{MR}) which only includes spatial integration. For $N=0$ this formula simplifies to
\begin{equation}\label{S}
s_1^{\text{RW}} = \int_0^{\infty} dx \left[q(x,1)\ln P(x,1) - q(x,0)\ln P(x,0)\right].
\end{equation}

After cancelations we obtain
\begin{equation}\label{actionRW2}
    s_1^{\text{RW}}=-2 n_0 \int_{0}^{\infty} d\mu\,\ln \text{erf}\,\mu = -\Lambda_1 n_0,
\end{equation}
where $\Lambda_1=2.06883 \ldots$. As a result,
\begin{equation}\label{1dquenched}
-\ln {\mathcal P}^{\text{RW}}\simeq \Lambda_1 n_0 \sqrt{T}
\end{equation}
has the same scaling
with $n_0$ and $T$ as in Eq.~(\ref{survivaldecay1}).
The coefficient $\Lambda_1$, however, is different from the coefficient $\Lambda_0=2/\sqrt{\pi} = 1.128379\ldots$ obtained for the random initial condition (see Ref. \cite{BB03}, Sec.~\ref{app:annealed} and Appendix). Equation~(\ref{1dquenched}) can also be deduced from an exact microscopic derivation when the particles are initially distributed periodically in space, see Ref. \cite{RM} and Appendix. The final density distribution~(\ref{qt1}) can  also be obtained from microscopic arguments.

\subsubsection{Finite-density correction}
\label{weakly}

Now let us go back to the SSEP and consider a small but finite $n_0$. We can calculate a small correction $\delta s$ to the action (\ref{actionRW2}) by treating the term  $h_1=-q^2 (\partial_x p)^2$ of the SSEP Hamiltonian (\ref{w}) perturbatively. In the first order of perturbation theory we have
\begin{eqnarray}
  \delta s &=&-\int_0^1 dt \int_0^{\infty} dx\, h_1[q_0(x,t), p_0(x,t)] \nonumber \\
  &=& \int_0^1 dt \int_0^{\infty} dx\,q_0^2(x,t) v_0^2(x,t), \label{s1a}
\end{eqnarray}
where $q_0(x,t)$ and $v_0(x,t)$ are the \emph{unperturbed} solutions, given by the RW formulas (\ref{qRW}) and (\ref{vRW}).
Plugging Eqs.~(\ref{qRW}) and (\ref{vRW}) into Eq.~(\ref{s1c}) we obtain
\begin{eqnarray}
  \delta s&=& \frac{2n_0^2}{\pi^2}  \int_0^1 dt \int_0^{\infty} dX\,\frac{e^{-\frac{2 X^2}{1-t}}}{t(1-t)}\,\nonumber \\
  &\times& \left[\int_0^{\infty} d\mu\, \frac{e^{-\frac{(X-\mu )^2}{t}}-e^{-\frac{(X+\mu)^2}{
   t}}}{\text{erf} \,\mu}\right]^2,
 \label{s1c}
\end{eqnarray}
where $X=x/2$.
To evaluate the above integral, we first replace the square of the integral over $\mu$ by a product
of two identical integrals over $\mu_1$ and $\mu_2$. The integration over $X$ reduces to calculating Gaussian integrals:
\begin{eqnarray}
  && \!\!\!\!\!\!\!\!\!\!\int_0^{\infty} dX\, e^{-\frac{2 X^2}{1-t}}
   \left[e^{-\frac{\left(X-\mu_1\right){}^2}{t}}-e^{-\frac{\left(X+\mu_1\right){}^2}{t}}\right] \nonumber \\
   &\times& \!\!\!\left[e^{-\frac{\left(X-\mu_2\right){}^2}{t}}-e^{-\frac{\left(X+\mu_2\right){}^2}{t}}\right] \nonumber \\
   &=& \!\!\!\sqrt{\frac{\pi\, t(1-t) }{2}}\,e^{-\frac{\left(\mu_1+\mu_2\right){}^2 (t+1)}{2 t}}\left(e^{\frac{2 \mu_1 \mu_2}{t}}-e^{2 \mu_1 \mu_2}\right). \label{intX}
\end{eqnarray}
Now we perform integration over $t$ in Eq.~(\ref{s1c}):
\begin{eqnarray}
   && \int_0^1 dt\, \sqrt{\frac{\pi}{2\, t(1-t)}}\,e^{-\frac{\left(\mu_1+\mu_2\right){}^2 (t+1)}{2 t}}\left(e^{\frac{2 \mu_1 \mu_2}{t}}-e^{2 \mu_1 \mu_2}\right) \nonumber\\
  &=& \frac{\pi^{3/2}}{\sqrt{2}} \left[e^{-\frac{1}{2} \left(\mu_1+\mu_2\right){}^2}
   \text{erfc}\left(\frac{|\mu_1-\mu_2|}{\sqrt{2}}\right) \right. \nonumber \\
   &-& \left. e^{-\frac{1}{2} \left(\mu_1-\mu_2\right){}^2}
   \text{erfc}\left(\frac{\mu _1+\mu _2}{\sqrt{2}}\right)\right] \equiv I(\mu_1, \mu_2), \label{intt}
\end{eqnarray}
where $\text{erfc} \,z=1-\text{erf}\,z$. The remaining double integral over $\mu_1$ and $\mu_2$ is evaluated numerically to yield
\begin{equation}\label{intmu}
\delta s=\frac{2 n_0^2}{\pi^2}\int_0^{\infty} \int_0^{\infty}  d\mu_1\,d\mu_2\, \frac{I(\mu_1, \mu_2)}{\text{erf} \, \mu_1\,\text{erf} \,\mu_2}
= \Lambda_2 n_0^2,
\end{equation}
where $\Lambda_2=1.08337\ldots$.  Therefore,
\begin{equation}
-\ln {\mathcal P} = \sqrt{T}\, s_1(n_0), \;\;\; s_1(n_0) = \Lambda_1 n_0 +  \Lambda_2 n_0^2 +\ldots. \label{actionweak}
\end{equation}

\subsubsection{Numerical solution}
\label{numeric1d}

We solved the MFT equations using a modification of the iteration algorithm, originally developed by Chernykh and Stepanov
\cite{Chernykh} for evaluating the probability density of large negative velocity gradients in the Burgers turbulence. Variants of this algorithm have been used in the context of MFT of lattice gases, with and without on-site reactions \cite{EK,MS2011,KM_var,KMS,MVS}. The algorithm iterates the diffusion-type equation~(\ref{d1})
forward in time from $t=0$ to $t=1$, and the anti-diffusion-type equation~(\ref{d2})
backward in time from $t=1$ to $t=0$. As in Ref. \cite{MVS}, our implementation of this algorithm involved an implicit finite difference scheme, which is beneficial for iteration convergence. At fixed $n_0$ and $\lambda$ we continued iterations until local convergence of the solutions was achieved with a high accuracy. Then we increased $\lambda$ and repeated the solution until the action (\ref{actionscaled2}) converged to 1 per cent. We also verified that,
for large $\lambda$ that we achieved, the mass loss to the absorber was negligible.

\begin{figure}
\includegraphics[width=0.35\textwidth,clip=]{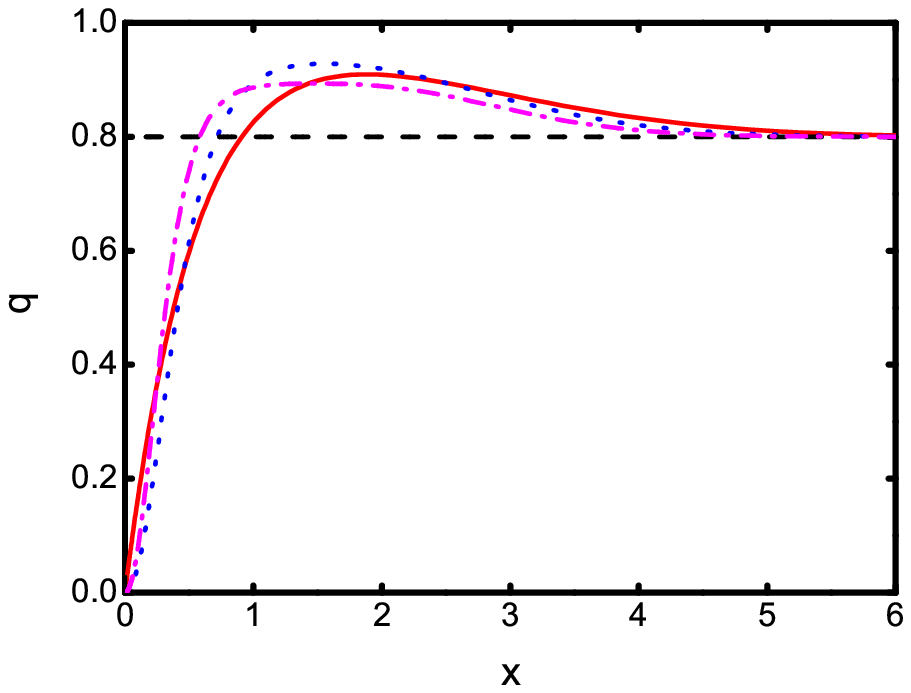}
\includegraphics[width=0.35\textwidth,clip=]{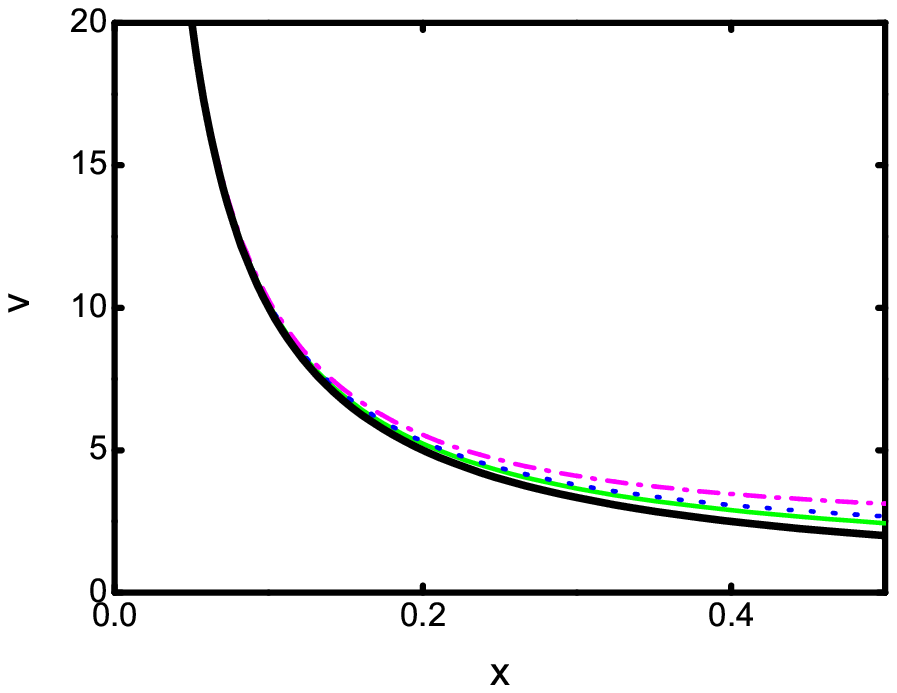}
\caption{(Color online) Numerically computed most likely density history of the SSEP for $d=1$, conditional on the zero flux to the target (located at $x=0$) for a deterministic initial condition with density $n_0=0.8$. Upper panel: $q$ versus $x$ at $t=0$ (dashed line), $0.25$ (dash-dotted line), $0.5$ (dotted line) and $1$ (solid line). Lower panel: numerically computed $v=\partial_x p$ versus $x$ at $t=0.25$ (dash-dotted line), $0.5$ (dotted line) and $0.75$ (thin solid line). The thick solid line shows the universal asymptotic $v=1/x$. The coordinate $x$ is rescaled by $\sqrt{T}$ (here $D_0=1$), and time is rescaled by $T$.}
\label{SSEPnum}
\end{figure}

Figure~\ref{SSEPnum} shows an example of our numerical solution for the deterministic initial condition and $d=1$. At small and moderately large $n_0$, the density history of the system is similar to that for RWs, with a rapidly forming density void
accompanied by a density peak. The density peak is lower than for the RWs, and it becomes progressively lower and broader as $n_0$ approaches $1$.  The numerically found $v(x,t)=\partial_x p(x,t)$ exhibits, at small $x$, the universal asymptotic (\ref{sing1d}).

\begin{figure}
\includegraphics[width=0.48\textwidth,clip=]{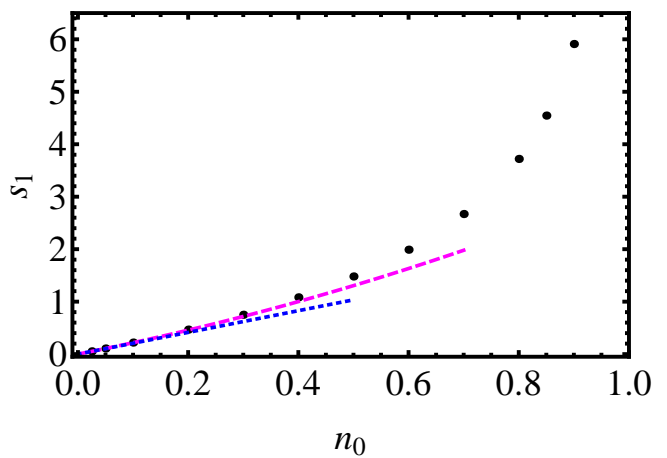}
\caption{(Color online) The function $s_1(n_0)$ found numerically for the deterministic initial condition. Shown are numerical data (points), the RW
asymptotic (\ref{actionRW2}) (the dotted line) and the finite-density asymptotic (\ref{actionweak})
(the dashed line).}
\label{snumeric}
\end{figure}

Figure~\ref{snumeric} shows the numerically found $s_1(n_0)$ for the deterministic initial condition and $d=1$. For small $n_0$, there is an excellent agreement with the RW asymptotic (\ref{actionRW2}). For moderate $n_0$, the results agree with the weakly-nonlinear asymptotic (\ref{actionweak}). As $n_0$ continues to grow, $s_1$ grows more rapidly. It must diverge at $n_0=1$, because in this case $-\ln {\mathcal P} (T)$ scales with time as $T$ rather than $\sqrt{T}$, as follows from simple microscopic arguments. Our numerical solution becomes prohibitive at $n_0$ very close to 1. The available data indicate the $(1-n_0)^{-1/2}$  divergence of $s_1$ as $n_0\to 1$.

A spherically symmetric three-dimensional version of the iteration algorithm was used for the verification of the stationary solution for $d=3$, presented in Sec. \ref{larged}.

\subsection{$d>1$, deterministic initial condition}
\label{nonstat3}

When $\ell\gg 1$, Eq.~(\ref{actionscaled2}) simplifies to
$$
s \simeq \Omega_d\,\ell^{d-1} \, \int_0^1 dt \int_{\ell}^{\infty} dr\, q(1-q)\, (\partial_r p)^2.
$$
The remaining double integral is equal to the rescaled action $s_1(n_0)$ in the (rescaled) \emph{one-dimensional} problem, $r\equiv x$, with an absorber at $x=\ell$. Because of the
translational invariance, $s_1=s_1(n_0)$ is independent of $\ell$.

As a result,
\begin{eqnarray}
  s &\simeq &\Omega_d\,\ell^{d-1} s_1(n_0)\label{sfast} \\
 -\ln {\mathcal P} &\simeq & S\simeq \Omega_d\,s_1(n_0) R^{d-1} \sqrt{T}. \label{Sfast}
\end{eqnarray}
Here $\ln {\mathcal P} $ is proportional to $\sqrt{T}$, rather than $T$.  In particular, for $d=3$
\begin{eqnarray}
  s &\simeq&  4 \pi \ell^2 s_1(n_0) \nonumber \\
 -\ln {\mathcal P}&\simeq& S\simeq 4 \pi s_1(n_0) R^2 \sqrt{T}.
 \label{3dshorttime}
\end{eqnarray}
The case of $d=2$ is not special here, and Eq.~(\ref{Sfast}) holds:
\begin{equation}
 -\ln {\mathcal P} \simeq \frac{8 \sqrt{\pi}}{3} s_1(n_0)\, R\sqrt{T}\quad\text{when}\quad R\gg\sqrt{T}.
 \label{nospecial}
\end{equation}

\subsection{$d=1$, Random initial condition}
\label{app:annealed}

In the annealed setting, that we consider here,
one allows equilibrium fluctuations in the initial
condition and averages over them. In a stochastic realization
of the process, the initial density profile is chosen
from the equilibrium probability distribution corresponding
to density $n_0$. As a consequence, the most
likely initial density profile, conditional on the target survival until time $T$, is
different from the flat profile $q=n_0$. The ``cost" of optimal fluctuation now
includes the cost of creating the optimal initial density profile. Still, the total
cost is less than the cost for the quenched (deterministic) initial condition,
so the survival probability for the annealed setting is higher than for the quenched setting.

In the MFT formalism, the annealed setting is described, in one dimension, by the initial condition that involves a combination of $q(x,t=0)$ and $p(x,t=0)$
\cite{DG2009b}:
\begin{equation}\label{inann}
p(x,0)-2\int_{n_0}^{q(x,0)} dq_1\,\frac{D(q_1)}{\sigma(q_1)} =\lambda \,\theta(x).
\end{equation}
For the SSEP, $D=D_0=1$ and $\sigma(q)=2q(1-q)$,
this becomes
\begin{equation}\label{inannssep}
p(x,0)-\ln \frac{(1-n_0)q(x,0)}{n_0[1-q(x,0)]} =\lambda \,\theta(x).
\end{equation}
For the RWs, $D=D_0=1$ and $\sigma(q)=2q$, we have
\begin{equation}\label{inannrw}
p(x,0)-\ln \frac{q(x,0)}{n_0} =\lambda \,\theta(x).
\end{equation}
Equation~(\ref{inannrw}) replaces Eq.~(\ref{t01}) in Sec. \ref{1d}. When $q(x,t)$ and $p(x,t)$ are found,
one can evaluate
$$
-\ln {\mathcal P}\simeq \sqrt{T} (s_0+s_1).
$$
Here $s_1$ is the action given by Eq.~(\ref{actionscaled1d}) [but with a different $q(x,t)$, see below],
whereas $s_0$ is the cost of creating the optimal initial condition $q_0(x)$.  This cost is given by the Boltzmann-Gibbs formula  \cite{DG2009b,KM_var,KMS}. For the SSEP
\begin{equation}\label{freeenSSEP}
    s_0=\int_0^{\infty} dx\left\{\ln \frac{1-q_0(x)}{1-n_0} + q_0(x) \,\ln\frac{q_0(x)(1-n_0)}{n_0[1-q_0(x)]} \right\},
\end{equation}
whereas for the RWs
\begin{equation}
\label{freeenergy}
    s_0=\int_0^{\infty} dx\,\left[n_0-q_0(x)+q_0(x)\,\ln \frac{q_0(x)}{n_0} \right].
\end{equation}

\subsubsection{Low-density limit: Non-interacting random walkers}

For the RWs, the annealed problem can be solved via the Hopf-Cole transformation.
In the new variables $Q$ and $P$, the initial condition (\ref{inannrw}) yields:
\begin{equation}\label{Q0ann}
    Q(x>0,t=0)=n_0 e^{-\lambda}.
\end{equation}
Solving the diffusion equation (\ref{Qt}) with this initial condition and the boundary conditions
$Q(0,t)=0$ and $Q(\infty, t)=n_0 e^{-\lambda}$, we obtain
\begin{equation}\label{Qann}
     Q(x,t)=n_0 e^{-\lambda}\, \text{erf} \left(\frac{X}{\sqrt{t}}\right),
\end{equation}
where $X=x/2$ as before. Now, $P(x,t)$ is still described by Eq.~(\ref{stepP}). Therefore, we can calculate $q(x,t)=Q(x,t) P(x,t)$.
Sending $\lambda$ to infinity, we arrive at
\begin{equation}\label{qann}
    q(x,t)=n_0 \,\text{erf}\left(\frac{X}{\sqrt{t}}\right)\,\text{erf}\left(\frac{X}{\sqrt{1-t}}\right),
\end{equation}
a symmetric function of $t-1/2$. There is no density peak in the annealed setting: the density is monotonically increasing with $x$ at all times. Interestingly, at $t=1$ and $t=0$ the optimal density
\begin{equation}
\label{q0ann}
q(x,1)=q(x,0)=n_0 \,\text{erf}\,X,
\end{equation}
is the same as predicted by the deterministic theory, Eq.~(\ref{difeq}), at $t=1$.  At times $0<t<1$, the optimal density profile $q(x,t)$ is a quadratic function of $x$ at small $x$ as before.

The action $s_1$ is given by Eq.~(\ref{actionRW1}) with the same $v(x,t)$ as in the quenched case, Eq.~(\ref{vRW}), and with $q(x,t)$ given by Eq.~(\ref{qann}). As in the quenched setting, it is more convenient to calculate $s_1$ using Eq.~(\ref{S}) that is equally valid in the annealed case. The cost of the initial condition $s_0$ can be evaluated from Eq.~(\ref{freeenergy}). Adding up $s_0$ and $s_1$, we obtain after cancelations
\begin{equation}
\label{s0s1}
    s_0+s_1=2n_0\int_0^{\infty}d\mu\, \text{erfc} \,\mu  = \frac{2n_0}{\sqrt{\pi}},
\end{equation}
so
\begin{equation}\label{annRW1}
 -\ln {\mathcal P}^{\text{RW}}\simeq \Lambda_0 n_0 \sqrt{T}
\end{equation}
with $\Lambda_0=2/\sqrt{\pi}$, in agreement with previous results \cite{BB03}, see also Appendix. To our
knowledge, the optimal density history (\ref{qann}) that contributes most to this survival probability,  has been previously unknown.

\subsubsection{Finite-density correction}

Now we return to the SSEP. Assuming $n_0\ll 1$, we can calculate a small correction ${\mathcal O}(n_0^2)$ to the expression $s_0+s_1$ from Eq.~(\ref{s0s1}).
The correction to $s_1$ is again calculated from Eq.~(\ref{s1a}), where $v_0(x,t)$ is still given by Eq.~(\ref{vRW}), but $q_0(x,t)$ is now given by the annealed history, Eq.~(\ref{qann}). We obtain
\begin{equation}\label{s1cann}
\delta s_1= \frac{n_0^2}{\pi}  \int_0^1 dt \int_0^{\infty} dx\,\frac{e^{-\frac{ x^2}{2(1-t)}}}{1-t}\,\text{erf}^2\left(\frac{x}{\sqrt{4t}}\right).
\end{equation}
The integral over $x$ can be evaluated using the formula 
\begin{equation*}
\int_0^{\infty} d\mu \, e^{-b \mu^2} \text{erf}^2 \mu = \frac{1}{\sqrt{\pi b}}\,
\arctan \frac{1}{\sqrt{b (b+2)}},\;\;b>0.
\end{equation*}
The remaining integral over $t$ is elementary,
\begin{equation*}
\int_0^1 \frac{dt}{\sqrt{1-t}}\,\arctan \frac{1-t}{\sqrt{4t}} = (3 - 2 \sqrt{2}) \pi,
\end{equation*}
and we obtain
\begin{equation}\label{deltas1}
    \delta s_1 = \frac{(3\sqrt{2}-4)n_0^2}{\sqrt{\pi}}.
\end{equation}
There is also a small correction to $s_0$ that comes from the difference of free energies of the SSEP and the RWs.
We calculate this correction by expanding the integrand of Eq.~(\ref{freeenSSEP}) in small $n_0$ and $q_0(x)$ up to, and including, the quadratic terms. The resulting correction is
\begin{equation}\label{deltas0}
   \delta s_0 = \frac{1}{2} \int_0^{\infty} dx\, [q_0(x)-n_0]^2= \frac{(2-\sqrt{2}) n_0^2}{\sqrt{\pi}},
\end{equation}
where we used the zero-order result (\ref{q0ann}) for $q_0(x)$. Adding up $\delta s_0$ and $\delta s_1$,
we finally obtain, for the annealed setting,
\begin{eqnarray}
  -\ln {\mathcal P}^{(\text{an})}  &\simeq & \sqrt{T}\, s_1^{(\text{an})}(n_0), \nonumber \\
  s_1^{(\text{an})} (n_0)&=& \frac{2}{\sqrt{\pi}} \left[n_0 +  (\sqrt{2}-1)\,n_0^2 +\ldots\right]. \label{actionweakan}
\end{eqnarray}
The $n_0^2$ correction agrees with the results of Santos and Sch\"{u}tz \cite{Santos}.  They solved a different problem for the SSEP, which involved particle injection  from the boundary into a semi-infinite line. Remarkably, that problem can be mapped, already at the exact microscopic level,  into the target survival problem we are dealing with here. As a result, the $n_0^2$ correction in the annealed setting, described by Eq.~(\ref{actionweakan}),
corresponds to the second cumulant of the statistics of the total number of injected particles at time $t=T$, when the system is empty at $t=0$ \cite{Gunter}.

Overall, Eqs.~(\ref{actionweak}) and (\ref{actionweakan}) show that, in one dimension, the survival probability exhibits different $n_0$-dependences in the quenched and annealed settings.

\subsection{$d>1$, Random initial condition}

When $\ell\gg 1$, the $1d$ results for the annealed setting represent an essential ``building block" in all dimensions $d>1$.  Here one obtains
\begin{eqnarray}
  s &\simeq &\Omega_d\,\ell^{d-1} s_1^{(\text{an})}(n_0),\label{sfastan} \\
 -\ln {\mathcal P} &\simeq & S\simeq \Omega_d\, s_1^{(\text{an})} (n_0) R^{d-1} \sqrt{T}. \label{Sfastan}
\end{eqnarray}
These equations resemble Eqs.~(\ref{sfast}) and (\ref{Sfast}), except that the rescaled one-dimensional action for the annealed setting $s_1^{(\text{an})}(n_0)$ is different from the corresponding quantity  $s_1(n_0)$ for the quenched setting. For very small densities $s_1^{(\text{an})}(n_0)$ is described by Eq.~(\ref{annRW1}). For small but finite densities it is given by Eq.~(\ref{actionweakan}). For arbitrary $n_0$, it can be found numerically.

\section{Extension to general interacting lattice gases}
\label{extension}

Importantly, the steady-state equation~(\ref{steadyeqgen})  can be solved analytically for general $D(q)$ and $\sigma(q)$, thus extending our long-time
results for $d\geq 2$ to a whole family of diffusive
gases of interacting particles. Indeed, by denoting
\begin{equation}
\label{uf}
u(r)=r^{d-1} \,\frac{dq(r)}{dr}\;\;\; \text{and}\;\;\; f(r)=\ln \frac{D[q(r)]}{\sqrt{\sigma[q(r)]}}
\end{equation}
we can recast Eq.~(\ref{steadyeqgen}) into a linear first order ordinary differential equation (ODE),
$$
\frac{du}{dr}+\frac{df}{dr}\,u=0,
$$
whose general solution is
\begin{equation}\label{u(r)}
u(r)=C \exp[-f(r)],
\end{equation}
where $C=\text{const}$. Using Eq.~(\ref{uf}), we obtain one more first-order ODE that can be easily integrated. Using the boundary conditions $q(\ell)=0$ and $q(\infty)=n_0$ to determine the two integration constants, we obtain the solution for $q(r)$ in implicit form:
\begin{equation}\label{qgeneral}
\frac{\int_0^q\frac{D(z)}{\sqrt{\sigma(z)}}\,dz}{\int_0^{n_0}\frac{D(z)}{\sqrt{\sigma(z)}}\,dz}=
1-\left(\frac{\ell}{r}\right)^{d-2},\;\;\;d>2.
\end{equation}
This solution exists for all lattice gases for which the integrals in  Eq.~(\ref{qgeneral}) are bounded.
This puts a limitation on the behaviors of $D(q)$ and $\sigma(q)$ at $q\to 0$. For
example, let $D(q\to 0)\sim q^{\alpha}$ and $\sigma(q\to 0)\sim q^{\beta}$. Then the integrals
converge at $q=0$ if and only if
\begin{equation}\label{crit}
2\alpha-\beta+2>0.
\end{equation}
For the SSEP and RWs one has $\alpha=0$ and $\beta=1$. Therefore, the condition (\ref{crit}) is satisfied,
and the solution (\ref{qgeneral}) exists. The condition (\ref{crit}) is also satisfied for a family of
repulsion processes \cite{KrRepulsion}.

When the solution (\ref{qgeneral}) exists, the action is bounded leading to a nonzero target survival probability. The rescaled action is the following:
\begin{eqnarray}
  s &=& \frac{1}{2}\,\Omega_d\, \int_{\ell}^{\infty} dr\,r^{d-1}\, \sigma\, v^2 \nonumber\\
  &=& \frac{1}{2}\,\Omega_d\,\int_{0}^{n_{0}}dq\,\frac{dr}{dq} r^{d-1} \sigma \, v^{2} \nonumber\\
  &=& \frac{1}{2} (d-2) \Omega_d\,\ell^{d-2}
  \left[\int_{0}^{n_{0}}\frac{D(q)}{\sqrt{\sigma(q)}}\,dq\right]^{2},
  \label{scaledgeneral}
\end{eqnarray}
where Eq.~(\ref{u(r)}) and the steady-state relation $v=(D/\sigma) (dq/dr)$ have been used in the last step.
As a result,
\begin{equation}
\label{Pgeneral}
    -\ln {\cal P}  \simeq 
    \frac{1}{2} (d-2) \Omega_d\,R^{d-2}\,T\,\left[\int_{0}^{n_{0}}\frac{D(q)}{\sqrt{\sigma(q)}}\,dq\right]^{2}.
\end{equation}
This closed-form result solves
the target survival problem for a broad class of diffusive lattice gases.  It has the same structure as Eq.~(\ref{survivaldecay3}) except the $n_0$-dependence which is model-specific. When specialized to the RW and SSEP, Eq.~(\ref{Pgeneral}) yields
Eqs.~(\ref{survivaldecay3}) and (\ref{Actresultd}), respectively.

For $d=2$ we obtain, with logarithmic accuracy, the long-time asymptotic
\begin{equation}
\label{Pgeneral2d}
    -\ln {\cal P}  \simeq
    \frac{\pi T}{\ln \frac{\sqrt{D_0T}}{R}}\,\left[\int_{0}^{n_{0}}\frac{D(q)}{\sqrt{\sigma(q)}}\,dq\right]^{2}.
\end{equation}

As an additional illustration of the general results (\ref{qgeneral}) and (\ref{Pgeneral}), we consider a family of zero range processes (ZRP). A ZRP describes interacting (but not excluding) random walkers on a lattice: A particle at site $i$ can hop to a neighboring site with a rate $\mathcal{R}(n_i)$ that only depends on the number of particles $n_i$ on the departure site $i$. Naturally, $\mathcal{R}(0)=0$. If $\mathcal{R}^{\prime}(n)>0$,
the ZRP is described at the macroscopic level by $D(q)=\mathcal{R}^{\prime}(q)$ and $\sigma(q)=2 \mathcal{R}(q)$, see e.g. Ref.~\cite{KMS}. Therefore, $D(q)/\sqrt{\sigma(q)} = \mathcal{R}^{\prime}(q)/\sqrt{2\mathcal{R}(q)}= (d/dq) \sqrt{2 \mathcal{R}(q)}$. Evaluating the integrals in Eq.~(\ref{qgeneral}), we obtain for the stationary density profile:
\begin{equation}
\label{qZRP}
\frac{\mathcal{R}(q)}{\mathcal{R}(n_0)}=
\left[1-\left(\frac{\ell}{r}\right)^{d-2}\right]^2.
\end{equation}
In its turn, Eq.~(\ref{Pgeneral}) yields the long-time asymptotic of the target survival probability for the ZRP:
\begin{equation}\label{ActZRP}
-\ln {\mathcal P}\simeq  (d-2)\, \Omega_d\,\mathcal{R} (n_0)\,R^{d-2} T,\;\;\;\; d>2.
\end{equation}

\section{Discussion}
\label{discussion}

In this work we evaluated the survival probability ${\mathcal P}(T)$ of a spherical target of radius $R$ in a gas of unbiased diffusive particles (``searchers"), with density $n_0$, that interact with each other via exclusion as described by the SSEP. We also determined the most likely particle density history  conditional on the target survival until time $T$.  The results depend on the dimension of space $d$ and on the basic rescaled parameter $\ell=R/\sqrt{D_0T}$. When $\ell$ is small and $d>2$, ${\mathcal P}(T)$ is mostly contributed to by an exact \emph{stationary} solution of the macroscopic fluctuation theory (MFT) that we obtained. For large $\ell$, and for any $\ell$ in one dimension, the relevant MFT solutions are non-stationary. In this case $\ln {\mathcal P}(T)$ scales differently with $T$, $R$, $d$ and $n_0$, and it also depends on whether the initial condition is deterministic or random. These effects (for large $\ell$, and for any $\ell$ in one dimension) are also observed in the absence of exclusion: for non-interacting random walkers.  In the special case of $\ell\ll 1$ and $d=2$  logarithmic corrections to ${\mathcal P}(T)$ appear. Table \ref{table1} can serve as a quick guide to our main results for the survival probability for the SSEP in different limits. The long-time asymptotics of  the survival probability for a whole class of interacting lattice gases for $d>2$ and $d=2$ are given by Eqs.~(\ref{Pgeneral}) and (\ref{Pgeneral2d}), respectively.

\begingroup
\squeezetable
\begin{table}[ht]
\begin{ruledtabular}
\begin{tabular}{l l l}
 \hspace{16pt}Dimension \hspace{25.3pt} & \hspace{24.5pt}$\ell \ll 1$ \hspace{27pt}  \hspace{30pt} $\ell\gg 1$ \hspace{29.3pt}\\
  \hline
  $d=1$ deterministic \hspace{11pt} & \hspace{40pt} Eq.~(\ref{Sfast}) with $d=1$ \hspace{42.5pt}  & \\ \hline
  $d=1$ random \hspace{29.9pt} & \hspace{40pt} Eq.~(\ref{Sfastan}) with $d=1$ \hspace{42.5pt}  & \\ \hline
  $d=2$ deterministic \hspace{11pt} & \hspace{20pt} Eq.~(\ref{action2}) \hspace{20pt} \hspace{10pt} Eq.~(\ref{Sfast}) with $d=2$ \hspace{0.6pt} \\
  $d=2$ random \hspace{29.9pt} & \hspace{20pt} Eq.~(\ref{action2}) \hspace{20pt} \hspace{10pt} Eq.~(\ref{Sfastan}) with $d=2$\hspace{3.3pt} \\
  $d>2$ deterministic \hspace{10.9pt} & \hspace{20pt} Eq.~(\ref{Actresultd}) \hspace{24pt} \hspace{10pt} Eq.~(\ref{Sfast})\hspace{43.0pt} \\
  $d>2$ random \hspace{29.8pt} & \hspace{20pt} Eq.~(\ref{Actresultd}) \hspace{24pt} \hspace{10pt} Eq.~(\ref{Sfastan}) \hspace{40.4pt}
\end{tabular}
\end{ruledtabular}
\caption{${\mathcal P}(T)$ for the SSEP in different limits}\label{table1}
\end{table}
\endgroup

The MFT provides a simple interpretation to the fact that, at $d>2$, $-\ln {\mathcal P}(T)$
scales as $T$ at very long times, whereas at $d=1$ it scales as $\sqrt{T}$.
The difference in behavior is related to the existence or non-existence of a proper stationary
solution of the MFT equations.

In summary, the macroscopic fluctuation theory opens new directions in the classical problem of target survival
probability.

\medskip

\section*{Acknowledgments}
We are grateful to  Davide Gabrielli for a helpful comment and to
Gunter Sch\"{u}tz for a useful discussion of Ref.~\cite{Santos} and its connection to the target survival probability of the SSEP in one dimension. Two of us (BM and PLK) thank the Galileo Galilei Institute for Theoretical Physics for the hospitality and the INFN for partial support during the completion of this work. This research was supported by grant No.\ 2012145 from the
United States--Israel Binational Science Foundation (BSF).

\appendix*
\section{Survival probability from microscopic theory of RWs}
\label{app:micro}

Here we present microscopic derivations of the target survival probability ${\mathcal P}(T)$ for the non-interacting RWs in one, two and three dimensions.
We verify that, for $d=1$, ${\mathcal P}(T)$  depends on the initial conditions (random or deterministic). We also show that, for $d=2$ and $3$, the leading term of the asymptotic of  ${\mathcal P}(T)$ is independent of the initial conditions. For $d=3$  we reproduce, both in microscopic calculations and in the MFT framework, an exact result \cite{BKZ86} for ${\mathcal P}(T)$ in the random setting. Finally, we derive, for $d=3$, a more accurate asymptotic of ${\mathcal P}(T)$ for the deterministic setting. In all these calculations we set $D_0=1$.

\subsection{$d=1$}

We start with random initial conditions and first consider RWs on a large but finite interval $(0,L)$. A RW starts in the interval $(x,x+dx)$, where $0<x<L$, with probability $dx/L$. The probability that this RW does not hit the target (which is at the origin) until time $T$ is $\text{erf}\, (x/\sqrt{4T})$. Averaging over random particle locations at $t=0$ we obtain the average single-particle non-hitting probability
\begin{equation}
\int_0^L \frac{dx}{L}\,\text{erf}\!\left(\frac{x}{\sqrt{4T}}\right)
= 1 - \frac{1}{L}\int_0^L dx\,\text{erfc}\!\left(\frac{x}{\sqrt{4T}}\right).
\label{A1}
\end{equation}
For sufficiently large $L\gg \sqrt{T}$ we can assume that the number of RWs on the interval is equal to $n_0 L$.  Since all the $n_0L$ particles are independent, the probability that none of them hits the target is
\begin{equation*}
\left[1 - \frac{1}{L}\int_0^L dx\,\text{erfc}\!\left(\frac{x}{\sqrt{4T}}\right)\right]^{n_0L},
\end{equation*}
which in the $L\to\infty$ limit becomes
\begin{eqnarray}
{\mathcal P}^{\text{RW}}(T)&=&\exp\!\left[-n_0\int_0^\infty dx\,\text{erfc}\!\left(\frac{x}{\sqrt{4T}}\right)\right] \nonumber \\
&=& \exp\!\left(-\frac{2  n_0 \sqrt{T}}{\sqrt{\pi}}\right),
\label{anmicro}
\end{eqnarray}
in agreement with Eq.~\eqref{annRW1} and Ref. \cite{BB03}.

Now let the initial positions of our RWs be deterministic. One example of deterministic setting is a periodic one, with exactly one particle on each site $k=1,2, \ldots$ of a lattice with lattice constant $1/n_0$.
The probability that the particles that is initially located at site $k$s does not hit the target until time $T$ is $\text{erf}\, [k/(n_0\sqrt{4T})]$. The probability that neither of the particles hit the target is
\begin{equation}
{\mathcal P}^{\text{RW}}(T)=\prod_{k=1}^\infty \text{erf}\!\left(\frac{k}{n_0\sqrt{4T}}\right)
\end{equation}
Taking the logarithm we obtain
\begin{equation}
\ln {\mathcal P}^{\text{RW}}(T)=\sum_{k=1}^{\infty} \ln  \text{erf}\!\left(\frac{k}{n_0\sqrt{4T}}\right).
\label{summation}
\end{equation}
We are interested in the regime of $n_0\sqrt{T}\gg 1$, when the characteristic diffusion length is much larger
than the lattice constant.  The leading-order result can be obtained by replacing the summation in Eq.~(\ref{summation}) by integration. Here we present a more accurate result that also includes a pre-exponential factor. We use the asymptotic \cite{RM}:
\begin{equation}
\sum_{k=1}^{\infty} \ln
\text{erf}\,(k u)  \simeq -\frac{\Lambda_1}{2u} -\ln \sqrt{u} +\ln\, \pi^{3/4},\nonumber\;\;\;0< u\ll 1.
\end{equation}
Here $\Lambda_1=2 \int_{0}^{\infty} d\mu\,\ln \text{erf}\,\mu =2.06883 \ldots $, see Eq.~(\ref{actionRW2}). As a result,
\begin{equation}
\label{RW:periodic}
{\mathcal P}^{\text{RW}}(T)\simeq \sqrt{2} \,\pi^{3/4} n_0^{1/2} T^{1/4}\,\exp\!\left(-\Lambda_1 n_0 \sqrt{T}\right).
\end{equation}
The exponential factor is independent of details of the deterministic initial condition. It coincides with our MFT result (\ref{1dquenched}) and differs from the annealed result, Eqs. (\ref{anmicro}) and \eqref{annRW1} and Ref. \cite{BB03}.

The pre-exponential factor is non-universal: it depends on details of
the initial condition.  This dependence is quite sensitive, as can be seen if we change the periodic arrangement of RWs
at $t=0$ by putting exactly 2 particles on each \emph{even} lattice cite $2k$, $k=1,2,\ldots$ of the same lattice as before, leaving all odd sites empty. Repeating the calculations, we arrive at
\begin{equation}
\label{RW:periodic2}
{\mathcal P}^{\text{RW}}(T)\simeq \pi^{3/2} n_0 \sqrt{T}\,\exp\!\left(-\Lambda_1 n_0 \sqrt{T}\right),
\end{equation}
with the same exponent as in Eq.~(\ref{RW:periodic}) but a much larger pre-exponent.

\subsection{$d=2$}

In two dimensions, the probability $P(r,T| R)$ that a RW starting at the radial coordinate $r>R$ does not hit the target by the time $T$ has a cumbersome exact expression. In the long time limit, $\ell \ll 1$, it suffices to use the following asymptotic that is valid with logarithmic accuracy (see e.g. \cite{SEP_source}):
\begin{equation}
\label{prob:2d}
P(r,T| R) \simeq 1 - \frac{1}{\ln\frac{4}{\ell^2}}\,\Gamma \left(0,\frac{r^2}{4T}\right).
\end{equation}
Here $\Gamma(a,z)=\int_z^{\infty} t^{a-1} e^{-t} dt$ is the incomplete gamma function.
We now employ the same line of reasoning as in one dimension. For the random initial condition we average the probability (\ref{prob:2d}) over the random locations of the particles in the annulus $R\leq r\leq L$ and obtain the average single-particle non-hitting probability
\begin{equation*}
1-\frac{1}{\ln\frac{4}{\ell^2}} \int_R^L \frac{2r dr}{L^2-R^2}\,\Gamma\!\left(0, \frac{r^2}{4T}\right).
\end{equation*}
When $L \gg \sqrt{T} \gg R$, the number of RWs in the annulus is approximately equal to $\pi n_0(L^2-R^2)$. Therefore, the probability that no RW hit the target is
\begin{eqnarray*}
&& {\mathcal P}^{\text{RW}}(T) \\
&&=\left[ 1- \frac{4 T}{(L^2-R^2)\, \ln\frac{4}{\ell^2}} \int_{\frac{\ell^2}{4}}^{\frac{L^2}{4T}}dz\,\Gamma(0,z)\right]^{\pi n_0(L^2-R^2)}\\
&&\to\exp\!\left[- \frac{4\pi n_0 T}{\ln\frac{4}{\ell^2}}\,\int_0^\infty dz\,\Gamma(0,z)\right],
\end{eqnarray*}
where we have simplified the limits of integration by recalling that $\ell\ll 1$ and taking the limit of $L\to\infty$. Computing the integral $\int_0^\infty dz\,\Gamma(0,z)=1$, we recover Eq.~(\ref{survivaldecay2}).

In the deterministic setting the probability is
\begin{equation}
\label{prob:2d-periodic}
{\mathcal P}^{\text{RW}}(T)=\prod_{r_j\geq R} \left[1 - \frac{1}{\ln\frac{4}{\ell^2}}\,\Gamma\!\left(0,\frac{r_j^2}{4T}\right)\right]
\end{equation}
The product is taken over initial positions $r_j$ which are deterministic. We assume that, at $t=0$, there is exactly one particle on each site of a square grid with lattice spacing $n_0^{-1/2}$ outside of the circular target of radius $R$.  We take the logarithm of \eqref{prob:2d-periodic} and, ignoring pre-exponential factors in the final result,  expand the logarithm to the leading order and replace the summation by integration by virtue of $\ell\ll 1$. We obtain
\begin{eqnarray*}
\ln {\mathcal P}^{\text{RW}}(T) &\simeq&
              -\frac{n_0}{\ln\frac{4}{\ell^2}} \int_R^{\infty} 2\pi r \, \Gamma\left(0,\frac{r^2}{4T}\right) \,dr\\
                                     &=&-\frac{4 \pi n_0 T}{\ln\frac{4}{\ell^2}}\,\int_{\ell^2/4}^\infty dz\, \Gamma (0,z)\\
                                     &\simeq& -\frac{4 \pi n_0 T}{\ln\frac{4}{\ell^2}},
\end{eqnarray*}
again arriving at Eq.~(\ref{survivaldecay2}). That is, in contrast to $d=1$, here the leading-order results  for
${\mathcal P}^{\text{RW}}(T)$ for random and deterministic initial conditions coincide in the limit of $\ell \ll 1$.

\subsection{$d=3$}

In three dimensions, the probability $P(r,T| R)$ that a RW starting at the radial coordinate $r>R$ does not hit the target until time $T$ can be found by solving the backward diffusion equation (which is mathematically identical to the forward diffusion equation)
\begin{equation*}
\frac{\partial }{\partial T}\,P(r,T| R)=
\left(\frac{\partial^2}{\partial r^2} +\frac{2}{r}\,\frac{\partial}{\partial r}\right)P(r,T| R)
\end{equation*}
subject to
\begin{equation*}
P(r,T=0| R)=1, \quad P(r=R,T>0| R)=0.
\end{equation*}
In contrast to two dimensions, the solution now has a simple form:
\begin{equation}
\label{prob:3d}
P(r,T| R)= 1 - \frac{R}{r}\,\text{erfc}\!\left(\frac{r-R}{\sqrt{4T}}\right).
\end{equation}
When the initial locations are random, we start with a spherical annulus $R\leq r\leq L$ and average \eqref{prob:3d} over initial locations to yield the average single-particle non-hitting probability
\begin{equation*}
1 - \frac{3R}{L^3-R^3}\int_R^L dr\, r\,\text{erfc}\!\left(\frac{r-R}{\sqrt{4T}}\right).
\end{equation*}
The probability ${\mathcal P}^{\text{RW}}(T)$ that no RW hit the target is given by
\begin{eqnarray}
\label{twoterms}
&& \!\!\!\!\!\!\ln {\mathcal P}^{\text{RW}}(T) \nonumber \\
&&\!\!\!\!\!\!=\frac{4\pi n_0(L^3-R^3)}{3} \ln \left[1-\frac{3R}{L^3-R^3}\int_R^L dr\,
r \,\text{erfc}\!\left(\frac{r-R}{\sqrt{4T}}\right)\right]\nonumber\\
&&\!\!\!\!\!\!\to- 4\pi n_0 R \sqrt{4T}\int_R^\infty dr\, r\,\text{erfc}\!\left(\frac{r-R}{\sqrt{4T}}\right)\nonumber\\
&&\!\!\!\!\!\!=- 4\pi n_0 R\, \sqrt{4T}\int_0^\infty dx\, (x \sqrt{4T}+R)\,\text{erfc}\, x \nonumber \\
&&\!\!\!\!\!\!= - 4\pi n_0 R\,T-8\sqrt{\pi}\, n_0 R^2 \sqrt{T},
\end{eqnarray}
in agreement with previous results \cite{BKZ86}. Equation~(\ref{twoterms}) is valid for any $\ell$. When
$\ell =R/\sqrt{T} \ll 1$, the first term is the leading one and yields Eq.~\eqref{survivaldecay3} with $d=3$. In the opposite case of $\ell \gg 1$ it is the second term that is the leading one, and it yields Eq.~(\ref{Sfastan}) with $d=3$.

Even in the long-time limit, $\ell =R/\sqrt{T} \ll 1$, Eq.~(\ref{twoterms}) is more accurate then the leading-order asymptotic~\eqref{survivaldecay3} that stems from the steady-state MFT solution.  Importantly, the final result~(\ref{twoterms}) can also be obtained from the MFT formalism if one solves the full time-dependent problem. The problem formulation is almost identical to that for $d=1$, see Sec.  \ref{app:annealed}, except that Eq.~(\ref{inann}) is replaced by
\begin{equation}\label{inannapp}
p(r,0)-2\int_{n_0}^{q(r,0)} dq_1\,\frac{D(q_1)}{\sigma(q_1)} =\lambda \,\theta(r-R),
\end{equation}
and all integrations over $x$ from $0$ to $\infty$ are replaced by integrations over the whole space outside the target. The calculations proceed along the lines of Sec. \ref{app:annealed}. The most likely gas density history,
in the original (not rescaled) variables, is
\begin{eqnarray}
  q(r,t)&=& n_0 \left[1-\frac{R}{r}\,\text{erfc} \left(\frac{r-R}{\sqrt{4 t}}\right)\right] \nonumber \\
  &\times & \left\{1-\frac{R}{r}\,\text{erfc} \left[\frac{r-R}{\sqrt{4 (T-t)}}\right]\right\}. \label{qann3d}
\end{eqnarray}
The calculation of the target survival probability ultimately reduces to evaluating the same integral as in Eq.~(\ref{twoterms}),  giving the same result.  In the long-time limit, $\ell \ll 1$, this integral is mostly contributed by the region $r-R \lesssim \sqrt{T}$. In this region Eq.~(\ref{qann3d}) can be approximated, up to small corrections, by $n_0 (1-R/r)^2$, which coincides with the steady solution~(\ref{qdRW}) for $d=3$. The deviations from the steady-state solution 
are responsible for the second term
on the right-hand side of Eq.~(\ref{twoterms}), which is a subleading term in this limit.

In the deterministic case the microscopic calculation, similar to that in $2D$, boils down to evaluating the integral
\begin{equation}
\label{deter3d}
\frac{\ln {\mathcal P}^{\text{RW}}(T)}{4 \pi n_0} = \!\int_R^{\infty} \!\!dr\,r^2 \ln \! \left[1\!-\!\frac{R}{r}\,\text{erfc}\left(\!\frac{r-R}{\sqrt{4T}}\right)\right].
\end{equation}
Exactly the same expression follows from the MFT. For $\ell \ll 1$, we expand the logarithm to the second order and arrive at
\begin{equation}\label{deter3dexp}
\ln {\mathcal P}^{\text{RW}}(T) \simeq - 4 \pi n_0 R\,T - (4-\sqrt{2})\sqrt{\pi}\, n_0 R^2 \sqrt{T}.
\end{equation}
The leading term coincides with that for the annealed setting, Eq.~(\ref{twoterms}). The subleading term is different.

For $\ell \gg 1$, we can replace $r$ by $R$ everywhere in the integrand of Eq.~(\ref{deter3d}) except under the $\text{erfc}$, thus arriving at Eq.~(\ref{Sfast}) (but for the RWs) with $s_1(n_0) = \Lambda_1 n_0$. Here the initial conditions affect the leading-order result.


\begin{thebibliography} {99}
\bibitem{dimensions} In our formulas for the SSEP we put the lattice constant $a$ to unity.
To restore the $a$-dependence, one should replace the particle number density $n$ by $n a^d$.
\bibitem{footnoteD} When $D$ depends on $q$,  $\ell$ can be defined as $R/\sqrt{D(n_0) T}$.
\end{thebibliography}
\end{document}